\documentclass{article}
\usepackage{graphicx}
\graphicspath{ {./images/} }
\usepackage[utf8]{inputenc}

\title{Conceptual Diagrams in Quantum Mechanics}
\author{J.E. Horvath and R. Rosas Fernandes\\
Universidade de S\~ao Paulo, {IAG}, S\~ao Paulo,Brazil}

\begin{document}

\maketitle

\centerline{\it foton@iag.usp.br} 

\centerline{http://orcid.org/0000-0003-4089-3440}

\bigskip\bigskip
\abstract{Quantum Mechanics (QM) stands alone as a (very) successful physical theory, but  the meaning of its variables and the status of many quantities in the mathematical formalism is obscure. This unique situation prompted the need of an attribution of a physical meaning to the latter, a procedure known as {\it interpretation}. On the other hand, the study of QM is usually presented, even to future scientists, within the only framework developed by Bohr and the Copenhagen researchers, known as the Copenhagen interpretation. As a contribution to the understanding and teaching of Quantum Mechanics, aimed to a broader and deeper appreciation of its fundamentals, including contemplating alternatives and updated interpretations for physicists and philosophers interested in the study of exact sciences (through Ontology, Epistemology, Logic or the Theory of Knowledge), we present a set of Conceptual Diagrams elaborated and designed to expose and facilitate the visualization of elements intervening in any interpretation of Quantum Mechanics, and apply them to several well-developed cases of the latter.}

\bigskip
\noindent
Keywords: Quantum Mechanics, Diagrams, Philosophy of Science


\maketitle
\vfill\eject

\section{Introduction}\label{sec1}

``Hard'' sciences are commonly associated to Mathematics and formal schemes, but they also comprise many other cognitive elements in their fundamental constitution. For instance, the use of some sort of diagrams is not a novelty in Science. In fact, there are many types of graphic elements employed to visualize and understand scientific issues, widely employed for teaching/learning in many disciplines, but which are sometimes a constituent element of a discipline.

About this statement, in an article entitled {\it Multiplying Meaning – Visual and Verbal Semiotics in Scientific Text}, J.L.Lemke [1] forcefully argues for the existence and need of non-verbal resources in scientific matters as follows:

``When scientists think, talk, work, and teach (cf. [1,2]) they do not just use words; they gesture and move in imaginary visual spaces defined by graphical representations and simulations, which in turn have mathematical expressions that can also be integrated into speech. When scientists communicate in print they do not produce linear verbal text; they do not even limit their visual forms to the typographical. They do not present and organize information only verbally; they do not construct logical arguments in purely verbal form. They combine, interconnect, and integrate verbal text with mathematical expressions, quantitative graphs, information tables, {\bf abstract diagrams}, maps, drawings, photographs, and a host of unique specialized visual genres seen nowhere else.'' (our bold)

As it stands, Lemke's statement gives a very important status to the graphic elements in Science. Examples of the type of elements which have become commonplace in modern science are abundant. One of the most outstanding cases is Venn's diagrams in set theory [3], developed during the 19th century when the definition of the present division of disciplines mostly took place. Venn's diagrams are now a part of the "disciplinary matrix" discussed by Kuhn [4] and Set Theory would be unthinkable today without them. 

Another paradigmatic case, this time of a different type, is the use of Feynman's diagrams in Quantum Field Theory. Initially devised as a tracking tool for the elementary terms of the S-matrix, their meaning is believed to be much wider, and their use is so widespread that contemporary practitioners first proceed to draw the diagrams for a given problem, and only later formalize their mathematical expressions. In the words of Veltman and t'Hoft [5]``...diagrams form the basis from which everything must be derived''. This deep symbiosis illustrates colorfully Lemke's thoughts: Feynman diagrams may be said to have become part of the {\it logos}.

The issue of the graphic representations has been suggested to play even a bigger role, in the very definition of Science. Latour [6] has argued for the uniqueness of Science to be related to the use of graphical elements, which facilitate the inscriptions and give at once a mobile, immutable and yet changeable character (in Latour's own definitions and words), quickly emerging since the Scientific Revolution. Even without subscribing this interesting thesis, there are a number of contemporary studies that contain an insight on graphs as semiotic resources (for example, Airey [7]) and agree on their central, key role.

When dealing with a wide and controversial subject, these resources can prove to be particularly important. The subject of our attention, Quantum Mechanics (QM), is about to complete a century of existence and currently enjoys a very special status: on the one hand it is recognized as one of the greatest creations of humanity in its repeated attempts to understand the Universe in which we live, and all its predictions have been confirmed through experiments; but on the other hand, its theoretical conceptions differ so much from the preceding traditions that QM led to a controversy that is far from being resolved. In other words, however different their theoretical conceptions and interpretations may be, the predictions of QM have always been confirmed when applied in experimentally, sometimes in flagrant contrast with intuitive classical expectations (see below).

In the foundations of these controversies, we can identify several elements that contradict the usual way that physics dealt with the objects of the world so far. In fact, the nature of microphysical reality and the way in which we apprehend the microphysical world were much questioned and led to alternative formulations which kept almost all of the original formalism, but gave a whole different meaning to the mathematical and physical elements, hence they are known as QM interpretations.

The existence of different interpretations on the most varied issues is quite peculiar in contemporary physics. Classical physics has a unique and exclusive interpretation of its own, since its formalism is unambiguous to physical reality. It should be added that an objective realism is also assumed, in the sense of admitting that physical objects exist independently of the observer. However, this is not so with QM, even within the widely accepted Copenhagen interpretation and other works seeking for a consistent meaning of the quantum formalism (see, for instance, de la Pe\~na [8]). Because of the deep meaning of the elements discussed in them, sometimes these interpretations are perceived as too broad and difficult, leading both scientists and teachers/students to misleading and many conceptually blocked paths.

According to the multirepresentation view [7, 9], an attempt to improve QM understanding must go beyond the grammatical language, mathematics and Aristotelian logic. We believe that the use of abstract diagrams can be converted into an effective tool to understand its interpretations and clarify the interconnection of the elements constituting the whole theory.

Therefore, the goal of this article is to propose a discussion bringing to light the main interpretations of QM, through conceptual diagrams that facilitate the understanding for future physicists and philosophers dedicated to the study of the quantum world.This article is organized and presented as follows: In Section 2, the conception of classical physics is presented and contrasted with Bohr´s Quantum Mechanics and the fully developed conception of QM as proposed by the Copenhagen researchers, the current ``orthodox'' version, highlighting some of its most important points.Section 3 is dedicated to a discussion of the constituent elements of QM, followed in the next Section 4 by the introduction of Conceptual Diagrams designed explicitly to facilitate the understanding of each interpretation of QM. The Conclusions and recommendations for the possible use of diagrams in the study and teaching in the classroom are given in Section 5.

\section{Classical Physics and Quantum Mechanics}\label{sec2}
\subsection{Classical Physics vs. QM}

It is true that Classical Physics, starting with Mechanics as initially proposed by Sir Isaac Newton, has in modern times a unique and exclusive interpretation, attributed to its mathematical formalism. As stated above, Classical Physics also implicitly contemplates objective realism by admitting that physical objects exist independently of the observer and that all experiments will obtain the same results as long as the same conditions are observed, that is, whenever the conditions of the experiment are compatible and analogous, however, the same does not happen with QM. It is true that the state-of-the-art of Classical Physics is the result of a long history of debates on the nature of space, the character of "forces" and other issues, but these have been sorted out and there is little or no trouble today.

Unlike Classical Physics, the conception of QM as a physical theory led from the very beginning to a series of issues that are quite deep and unsolved. In fact we can identify several elements in it that contradict the usual way that Classical Physics deals with the objects of the world, and even the nature of QM's own formulation seems different. This led to a variety of ways of dealing with the nature of Reality in the microphysics realm, and also the way in which we apprehend the microphysical world have been and still are much questioned. While retaining almost all of the original formalism, these formulations give different meanings to the mathematical and physical elements of QM, which is why they are known as ``QM interpretations''.

Consequently, a presentation of the foundations of QM, as initially conceived and developed by Niels Bohr and the Copenhagen School, as well as their subsequent interpretations are extremely relevant for the training and updating of the future professional physicists and future philosophers as well.

\subsection{Quantum Mechanics and its Postulates (brief overview)}

In the early 20th century, the pioneers of QM faced the challenge of building a physical theory for the micro world that presented major conceptual and formal problems. Indeed, the notions deriving from classical physics were not sufficient for this task. Grammatical language, mathematics and Aristotelian logic were also suggested to be insufficient for understanding QM. ``Our words do not fit'', a famous quotation expressed by Heisenberg [10] about their use in QM illustrates this very keenly.

QM was initially developed by Niels Bohr and the so-called Copenhagen School and it must be considered that with its later definitive reformulation, QM was given a probabilistic nature, resulting in debatable and controversial issues (notably by Einstein, Schr{\"o}dinger and others) and prompting a continuous search for a more adequate interpretation. Because of the well-known excellent sources on QM (i.e. Ismael [11]) our presentation of the subject will be very brief and essential. 

This entire formulation is usually taught, generation after generation, without touching on the numerous problems that arise as a result of its further interpretations. Indeed, as a initial problem, we know that in any physical theory, the experimenter measures some quantity and compares it with the prediction. In QM, however, it is said that the experiment will measure predefined values (eigenvalues). In fact, in QM we are obliged to accept statements such as: ``If the system is in an eigenstate of its observable $A$, corresponding to the eigenvalues, an observer measuring $A$ will certainly obtain the value a''. Objectively seen, the prescriptive dogmatic character of this type of framework is overwhelming, but it is presented as ``natural'' and inherent to QM without further discussion about it (at least within the orthodox Copenhagen interpretation), leaving much to be desired for those who need or want to go deeper into the issue [12].

The type of information desired for the quantum description required the formulation of the n-dimensional space of states (also called {\it Hilbert space}). It is postulated that all possible system states are contained in the state vector $\mid\Psi\rangle$ representing the system.

For each measurable quantity there is an Hermitian operator $\hat{Q}$, the mathematical character of these operators that act on the states guarantees the consistency of the results (for example, they eliminate imaginary probabilities).

The results of the physical measurements correspond to the $q$ eigenvalues of the operator $\hat{Q}$ with probabilities computed using the inner product in the Hilbert space. The dynamical evolution of the wavefunction/state vector $\mid\Psi\rangle$ in the so-called Schr\"odinger picture is 

\begin{equation}
    \hat{H} \mid\Psi\rangle = i \hbar \frac{\partial}{\partial t} \mid\Psi\rangle
\end{equation}

Once the solutions of eq.(1) are found,$\mid\Psi\rangle$ can be decomposed in a complete mathematical basis, formed by the solutions, and its temporal evolution 
is just

\begin{equation}
 \mid\Psi\rangle = \sum_{n} a_{n} \mid\Psi_{n}\rangle \exp{\bigl( -iE_{n} t/\hbar \bigr)}
\end{equation}

where $E_{n}$ are the eigenvalues of the Hamiltonian operator. According to the (postulated) structure pointed out above, a single measurement can only give as a result one of the eigenvalues of the system.
		
With the probabilistic character of the description, quantum phenomena are the true objects of description of the theory, independently of the existence of a quantum Reality (see below). In fact, a central foundation of Copenhagen QM version is that, if a measurement is taken, the wavefunction  $\mid\Psi\rangle$ {\it collapses} as a result of system-device interaction. There is a quantum ``leap'' that is not accessible to human understanding, and is not described by the formalism. It is further postulated that the measurement results are expressible only in classical terms. Niels Bohr insisted a lot on this point (and gave it the name of {\it correspondence}), for him the measurement results would not make sense without the existence of Classical Physics.

Another fundamental difference of QM with other theories results from the fact that operators do not always commute, and although much has been discussed about the experimental situation related to statistical dispersion, this property stems directly from the mathematical structure of the Hilbert space. The {\it commutator} of two operators $[\hat{p}, \hat{q}]$ called conjugate quantities is proportional to the reduced quantum of action. 

\begin{equation}
            [\hat{p}, \hat{q}] = i \hbar              .
\end{equation}                 

That is, the conjugate quantities simultaneously measured in QM cannot have definite values whose errors → 0 under any circumstances, as they emerge from a set of probabilities of eigenvalues of operators with an irreducible dispersion due to their quantum nature. The measurement process is considered the source of irreversibility, but it never enters into the calculation. Not even the observer “itself" nor the measuring apparatus appear anywhere in the formalism.

These features, among others, was never accepted by Einstein and other physicists, who supported the incomplete character of QM, that is, the determination of physical quantities with infinite precision when incorporated into a more comprehensive and complete theory. Criticisms of QM by Einstein, Schr\"odinger, and others referred to these obscure aspects of its formalism and also to the very idea of the quantum object. The famous case of Schrodinger's cat, for example, was enunciated through a {\it reductio ad absurdum} in which QM allows the display of a half-alive and half-dead cat simultaneously inside a closed box in which a poison bottle is activated through a quantum process. Schr\"odinger considered the quantum description of this situation as absurd. 

Over time these criticisms led Bohr's conception to further extreme positions: Reality came to be considered ultimately “metaphysical”, in the sense that it does not have a demonstrable existence, but it was stated that it is even dangerous to think about it. The Copenhagen group formulated a kind of ``No-go'' interpretation about QM, completely differentiating it from all the preceding theories. Thus, the most obvious path resulted in the evolution of plain Idealism (subjectivism) into an Operationalism position (see below). In the latter philosophical approach, there is no concern for the nature of Reality and related problems. Late QM, in turn, has been reduced by some to the idea of a set of rules of calculations, and, without denying that there is a Reality, does not refer to it. Nothing could upset Einstein more than this {\it ab initio} resignation, since he strongly believed that Physics needs to address the behavior of real objects in a objective world.

\section{Elements constituting QM interpretations}\label{sec3}

We will show below that through abstract diagrams it will be possible to distinguish all the elements that enter the interpretation, and thus point out that in QM not only the nature and existence (Ontology) of a Quantum Reality is questioned, but also the ideas of ``observer'' and ``phenomenon'', which may be different in each case.

As a first example of this problem, it must be considered the very separation between the subject (observer $S$) who experiences the world and a Reality object ($R$) to be apprehended by the subject through the analysis and observation of phenomena, which is central to QM (and all physical theories in fact).

For Western science, in the classical realm, this separation is so evident that it is not even discussed or mentioned anywhere. However, for other ways of thinking (essentially in Eastern philosophies, but also many native Americans), this separation between subject $S$ and the Reality object $R$ being studied becomes impossible. We will see that some of the interpretations of QM identify this separation as a source of fundamental discrepancy between experiments and ``reasonable'' expectations.

The so-called interface (Epistemology) with quantum phenomena is many times pointed out as a possible source of problems in QM. In general, any physical theory needs at least three fundamental elements to deal with the object of study (system). These are:

(1) a {\it Logic} based on syllogisms or other tools. As with the concept of ``observer-subject'' and supposedly separable from the system, there is the implicit hypothesis that the Logic of the world is Boolean (Aristotelian). Von Neumann was one of those who insisted on the possible non-human Logic of QM, in parallel with the case of non-Euclidean geometries. For example, in the class of statements like ``if {\it p}, then {\it p} or {\it q}'' quantum version ``if Schr\"odinger's equation is valid, then the system evolves according to it and a measurement will give one of the eigenvalues'' is constantly formulated without its listeners noticing its inconsistency within a Boolean logic.

(2) a consistent {\it Algebra} to manipulate basic objects ($\mid\Psi\rangle$, etc.) and obtain quantitative predictions. This is of course different from the underlying Logic in general, and is fundamental when obtaining concrete results (for example, the inner product in Hilbert space allows to calculate the probabilities of each eigenvalue in a single measurement).

(3) Finally, a {\it Language} (formal or informal), which brings as a corollary a semantics-semiology, not always properly scrutinized. Besides the remarkable quotation of Heisenberg (in Heisenberg [10]) ``our words do not fit'', this issue within a wider context (in the sense of the adequacy of human languages to formulate scientific statements) has been explored by Wittgenstein [13, 14] and others, and remains a major issue in the philosophy of Science, taking a dramatic turn for the interpretations of QM.

Finally, and at the risk of sounding absurd, the very idea of a physical object to be studied is not guaranteed in QM. We have pointed out that in a late Copenhagen interpretation, Bohr even stated that QM is not about Reality, but about what can be said about phenomena [15], plainly stating that the late Copenhagen version of QM was an epistemological theory. Furthermore, in various versions of QM one can find a certain philosophical Idealism, namely that the physical world is actually a product of the mind. Thus, the question of the empirical content of QM takes dramatic dimensions, a question that should not be ignored.

\section{Diagrams and QM Interpretations}\label{sec4}

For a complete visualization of each interpretation of QM, a series of illustrative conceptual diagrams that explain the basic elements and their role in them are here created and presented.

In the diagrams, the Subject (observer) is represented in the diagrams with a triangle with the letter $S$, and it would be ultimately important to distinguish whether it has a ``consciousness'' or otherwise, for example, being a simple measuring device (although we will not attempt to develop this issue, well beyond the scope of this article).

 {\it Epistemology} (horizontal thick arrows) is the set of empirical (experiences) and formal (Algebra and Language) tools for an assumed Logic, with which it is intended to apprehend the quantum phenomena QP (represented by an asterisk), which in turn are supposed to be produced by a QR object(s), if existing (we stress again that some representations deny the very existence of a deeper quantum reality). The human physiological/linguistic limitation discussed by Wittgenstein [13, 14] and other authors gives rise to a buffer we called $W \, filter$, always explicitly indicated, that shapes and limits the subject's $S$ perception and understanding. We will call the procedures of this epistemological connection generically as measurements. A {\it phenomenon} is marked with an asterisk in all cases.

 Finally, the {\it Ontology} is represented by an ellipse that includes the existing objects according to each interpretation. This can be classical (having defined values of physical quantities at all times) or quantum (without defined values for any time) or displaying a non-classical feature (such as entanglement of phases), in which case we have written ``classical'' within quotation marks. Finally, quantum objects may be non-existing within the interpretation, or at least sometimes not having an explicit characterization.

\subsection{The Classical Physics case}

The Classical Physics case is a benchmark to grasp what a conceptual diagram can deliver. Few objections against the classical picture have been raised, and this is why there are no current discussion on the ``interpretations'' of Classical Physics (except for some specific issues). In terms of our definitions above, the diagram describing the classical situation (Fig. 1) can presented as follows

• The ``Reality'' (Classical Reality, or CR here), the objects and their properties, the measured phenomena (asterisk) and the Subject $S$ itself are causally separated, when measured, the properties are well defined for any time and are local (they do not depend on the distant environment, and they exist even if not measured by hypothesis). The phenomena are manifestations of existing CR object(s) and the task of Physics is to know the latter through the measurements and formulation of compelling theories. This complies with the Realism of A. Einstein and most physicists, although a group of putative Idealists may challenge the statements even within the classical realm.

• As stated, there is no need for any ``interpretation'' because classical theory defines objects with definite values of their physical quantities unambiguously for all times (hence the circle in ``CR'', which will be replaced by a ``cloud'' in ``QR.'' Wittgenstein's filter $W$ has here a secondary role, at least not as central as the one it will play in Quantum Mechanics.

\includegraphics[scale=0.4]{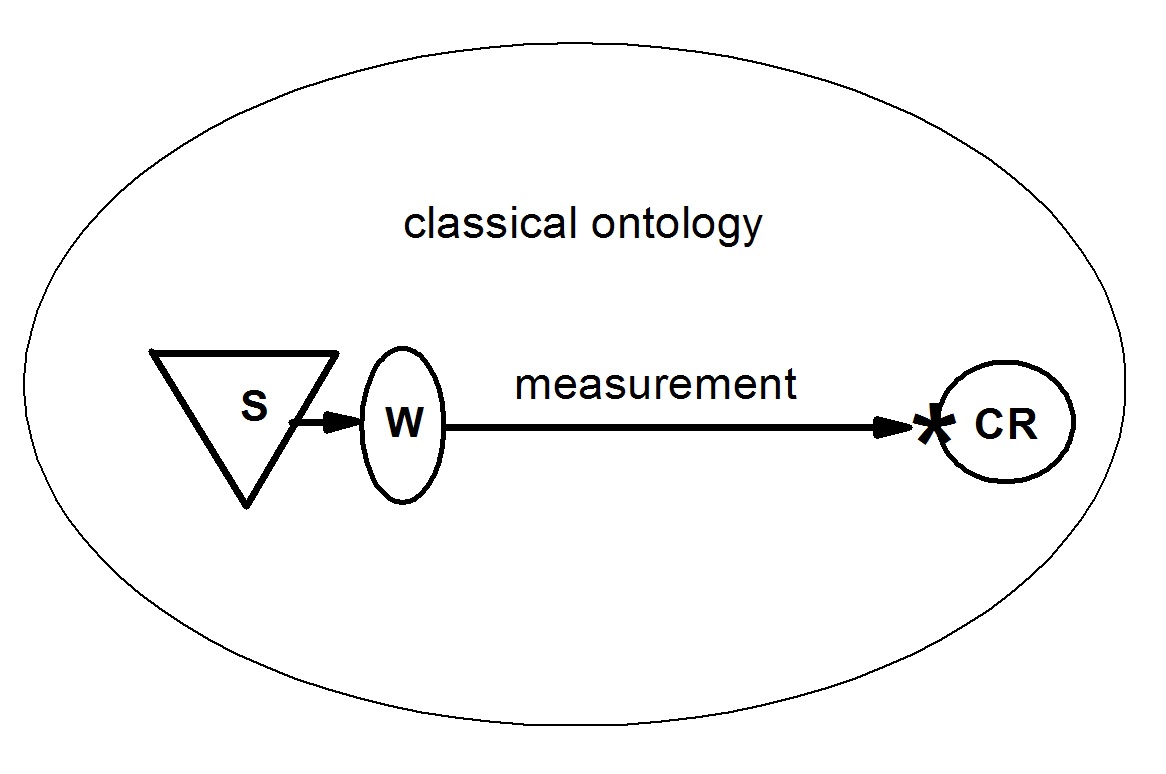}

Figure 1.The Classical Physics diagram, in which the elements are well-defined with little or no dispute.

\section{Conceptual diagrams for QM interpretations}

\subsection{The Copenhagen interpretation}

\includegraphics[scale=0.4]{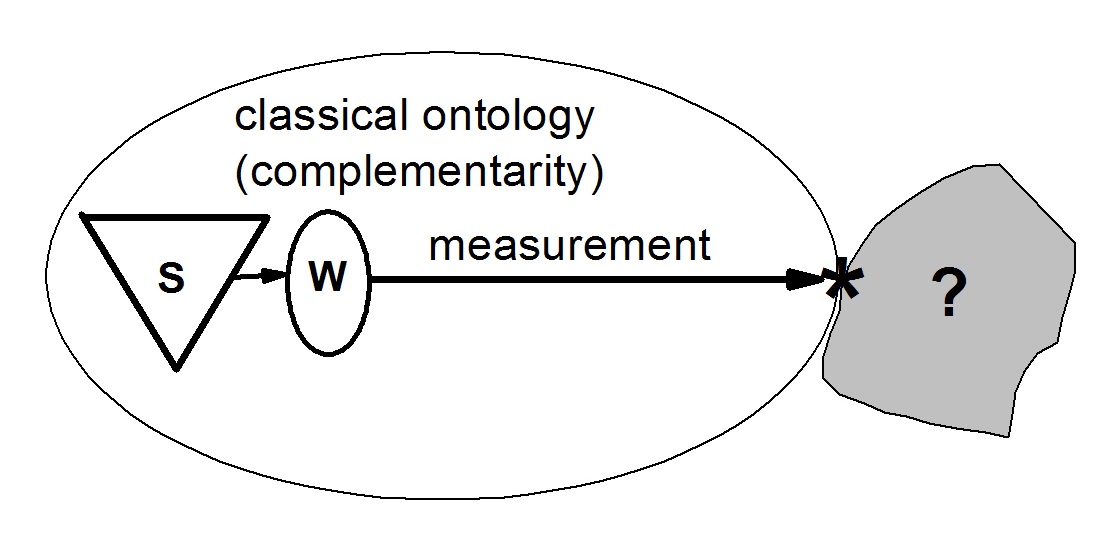}

Figure 2. The Copenhagen interpretation. The existence of a QR is denied (or at least not deemed necessary), although quantum phenomena exist and are the subject of QM. Their study, according to Bohr, would be impossible without the subject S, the measurement apparatus and the calculation rules belonging to the classical realm (Principle of Complementarity), all of them inside the large white ellipse.

As discussed above, the late statements by Bohr (see Petersen [15]) about the Copenhagen interpretation are truly remarkable: it is said that there is no concrete Quantum Reality, objects and their properties do not have defined values (they do not exist!) before being measured. The act of measuring is what "creates the Reality", that is, it defines the type of phenomenon measured. The subject, measuring devices and results are expressible only in classical terms (Bohr insisted that this complementarity is crucial to be able to say something about the quantum phenomena). It is granted by the Copenhagen interpretation that the Logic of QM is Aristotelian (Boolean), but the results show the probabilistic nature of the phenomenon, because this is all we can get for them as a result: a set of probabilities. QM is therefore a theory that says how much we can say about objects (epistemological) and does not describe any true Quantum Reality (these points were absurd to Einstein, who claimed that they were enough to think that QM is incomplete and would be surpassed by a better approach to a realistic physical world, as advocated by him).

These unusual features have been largely debated and rejected by many physicists and philosophers, and praised by the majority of the scientist's community. Among the former, Bunge [16] was emphatic to declare the Copenhagen Interpretation plainly false, and advocated its substitution by a form based on (non-classical) realism. In any case, and even in its earlier form in which a Reality was not denied, it is clear that the latter was never a main object of worry by their creators, and this justifies the question mark on the gray zone in Fig. 2.

\subsection{The non-local Reality (de Broglie-Bohm et al.)}

\includegraphics[scale=0.4]{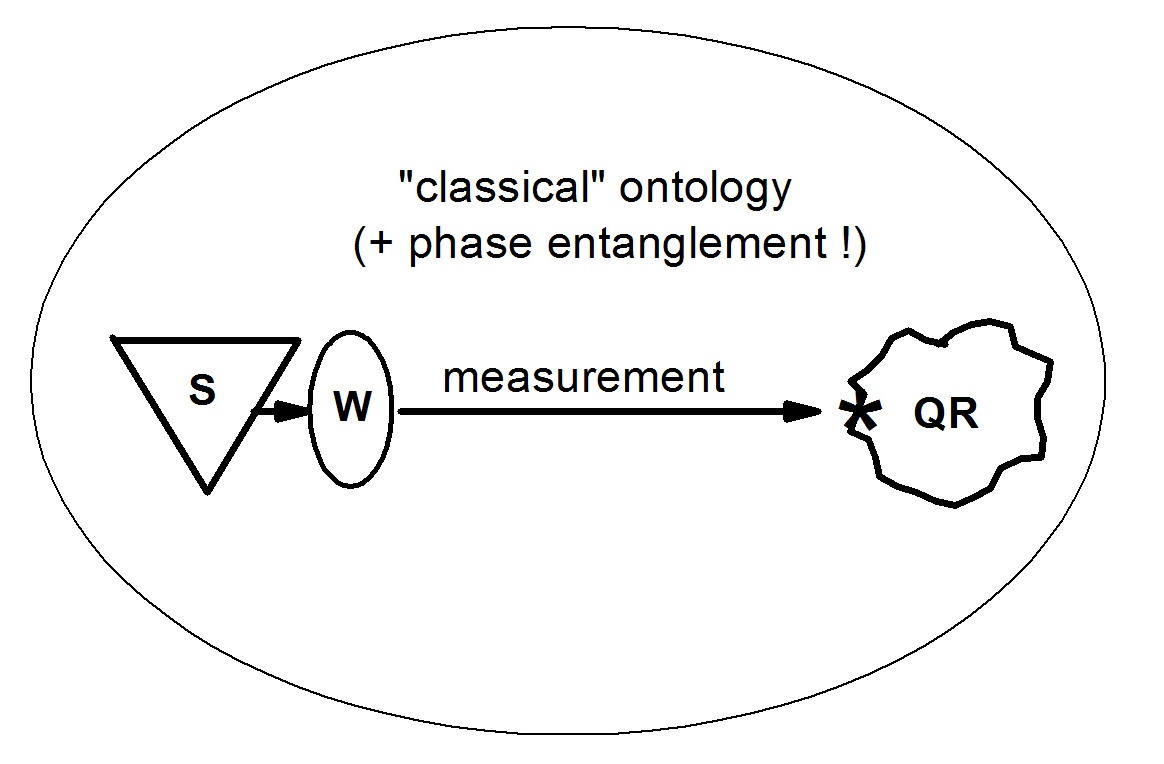}

Figure 3. The non-local interpretation of Bohm and de Broglie, supported by many experimental results. The main ingredient is the so-called phase entanglement of the wavefunctions, a feature that does not exist in Classical Physics, although ultimately the nature of the quantum objects is not that different from the classical ones.

In this version of the theory, it is postulated that the ``Quantum Reality'', objects and their properties, measured phenomena and the subject itself ($S$) cannot be separated. When measured, objects keep memory of their space and time history (a feature described as {\it phase entanglement}, Fig. 3). The entire Universe is an indivisible Whole (i.e., extremely non-local.) Particles ride the wave functions but remain hidden (hidden variables) without influencing them (it is not clear if they are ultimately superfluous). We have employed quotation marks for the word ``classical'' because entanglement is not a classical feature strictly speaking.

Around 50 years ago a battery of experimental tests were devised and performed to verify the non-locality of quantum objects (see a panoramic account in Herbert [17]). The initial results already pointed out that the agreement with the predictions of QM, and rejection of the locality of the QR was obtained. Since then, every experiment confirmed the non-locality and the idea that, to some extent at least, the idea Whole applies, and the locality is just an approximation valid within certain limits. It is impressive that this conclusion is not as widely known and discussed as it should, and attempts to save the local character have not came with a proper ``solution''. However, a claim that the de Broglie-Bohm interpretation of QM is fully confirmed by these experiments is premature. 

\subsection{The interpretation of the many-worlds (or {\it The Garden of the Forking Paths}, by J.L.Borges)}

\includegraphics[scale=0.4]{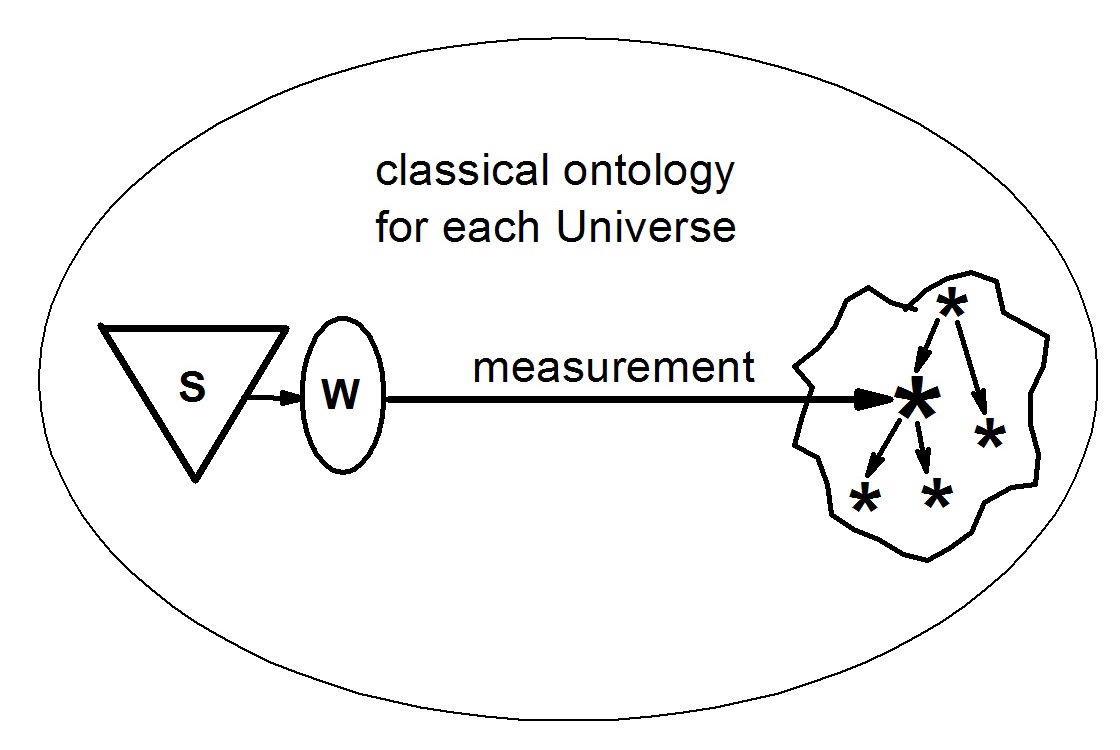}

Figure 4. Everett´s many-worlds interpretation, in which each measurement splits the future history.
 
This interpretation developed by H. Everett (see Pinto-Neto [18] for a discussion) holds that Quantum Reality is a set of systems disjoint in time. Quantum objects bifurcate their histories for each possible result of a measurement, and each version continue to exist in their own parallel Universes (which are real!, not a Gibbs ensemble of mental copies). Therefore, each measurement corresponds to one of the possibilities, and therefore there is no wavefunction collapse, but rather a probability of selecting one of the many components. The diagram of Fig. 4depicts symbolically this hypothesis and connects all the elements discussed in Section 4 to it.

It is obvious that the scenario leads to an amazing conception of the whole physical reality, at least from the philosophical point of view. A literature piece based on this idea (but independent of Everett's scientific work which is several years older) was written in the form of a short story by J.L. Borges [19], but without any explicit mention to QM, which was not a subject of the literary piece [20], nor appears anywhere in Borges' writings. Everett declared that he did not know Borges story either.

\subsection{Quantum Logic (Birkhoff-von Neumann)}

\includegraphics[scale=0.4]{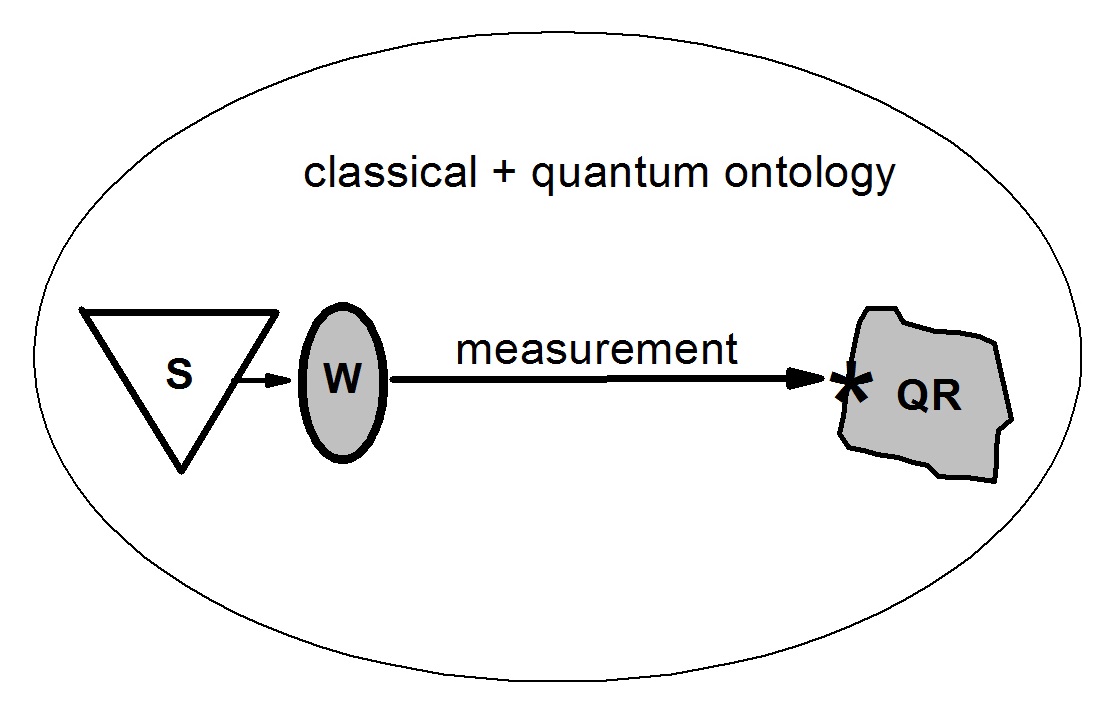}

Figure 5. The Quantum Logic of Birkhoff and von Neumann, in which the non-Boolean logic operating inside the W-filter is held responsible for the problems encountered in QM. The Logic of humans is not the Logic of Nature.

In the Birkhoff--von Neumann interpretation the main point is the examination of the subject $S$. According to them, observers are conditioned by the reasoning structure (Language), represented by the Boolean logic. Their postulate is simply that QM does not follow the latter, and a different type of logic is needed. Therefore, in this interpretation, the problems originate and restricted to what we have called the Wittgenstein filter that connects the measurement results with the subject-observer. This does not mean that a QR cannot exist, only that its apprehension is complicated by the inherent Logic which applies to the theory. Metaphorically we may state that QM talks to human observers in a foreign language, unknown to us, and we are compelled to discover and learn that language to understand the physical world (Fig.5).

\subsection{Consciousness creates Reality (von Neumann-Wigner-Stapp)}

\includegraphics[scale=0.35]{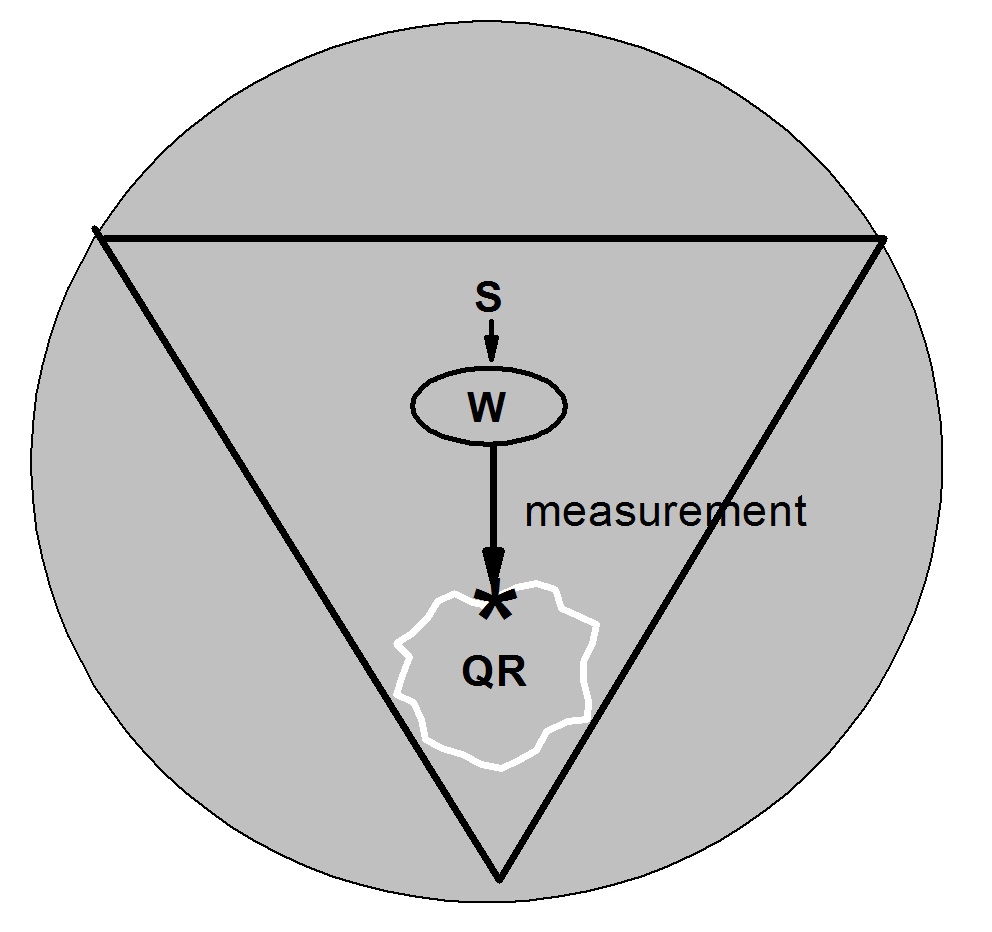}

Figure 6. Consciousness is the ultimate entity beyond QR, according to the Idealistic interpretation of von Neumann- Wigner- Stapp.

The interpretation of von Neumann-Wigner-Stapp is perhaps one of the most outrageous in all Physics, and raises controversies about the nature of reality and the existence of intelligence as well. It may be labeled in its final form as ``pure Berkeley Idealism'', because it suggests that consciousness creates Reality. It has a strong resonance in some oriental philosophies suggesting that the world is a kind of dream of a superior mind [21]. 

The interpretation is very polemic in itself, and was suggested by two outstanding contributors of Quantum Theory. Adopting it as true, it also beaks the remaining objectivity of orthodox QM in the sense that for the latter the measuring device could be an apparatus or a conscience, whereas in the von Neumann- Wigner- Stapp version only a conscientious observer stands, who not only measures, but also creates the observed phenomenon with its intervention. Quantum behavior is in this sense an attribute of Consciousness (Fig.6).

\subsection{Heisenberg's {\it potentia}}

\includegraphics[scale=0.45]{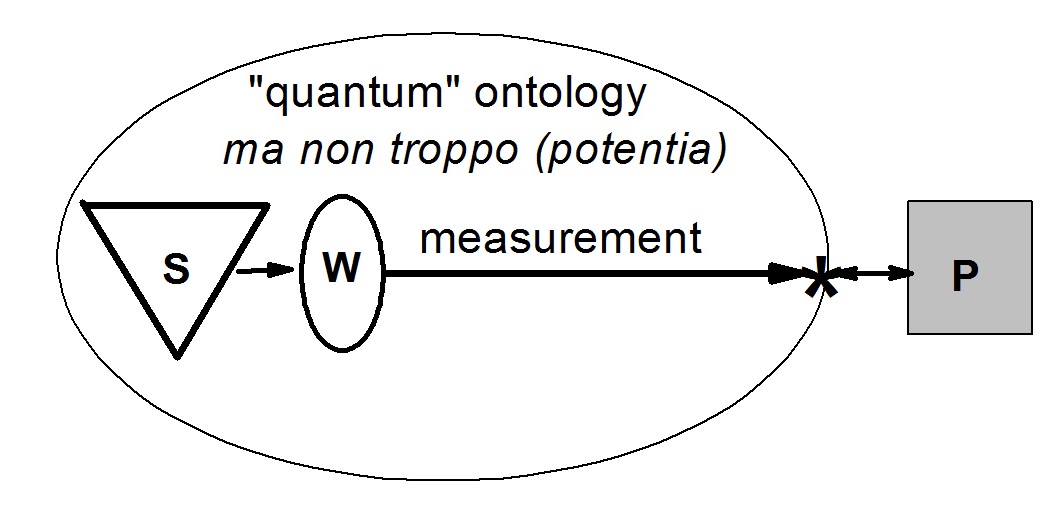}

Figure 7. Heisenberg´s proposal for the understanding of where the QP came from in the absence of a QR: the {\it potentia} concept. Note that we have not located a QR, from which the {\it potentia} P would extract the values in the diagram.

This interpretation is due to one of the founding fathers of QM, Werner Heisenberg, suggested in a later development many years after his own initial work and the discussion prompted by the QM in the Copenhagen interpretation. Heisenberg suggested that there is a kind of limbo (called {\it potentia} by him) in which objects of Quantum Reality exist, somewhat ``halfway'' between the QR and the observed quantum phenomena. The act of measuring defines the type of phenomenon, and the result stems from the world of {\it potentia} P, not from the Quantum Reality itself. The Fig. 7 attempts to depict Heisenberg's interpretation which introduces the {\it potentia} as a buffer of probabilities implied by the QM formalism.

\subsection{Statistical interpretation (Born-Einstein)}

\includegraphics[scale=0.4]{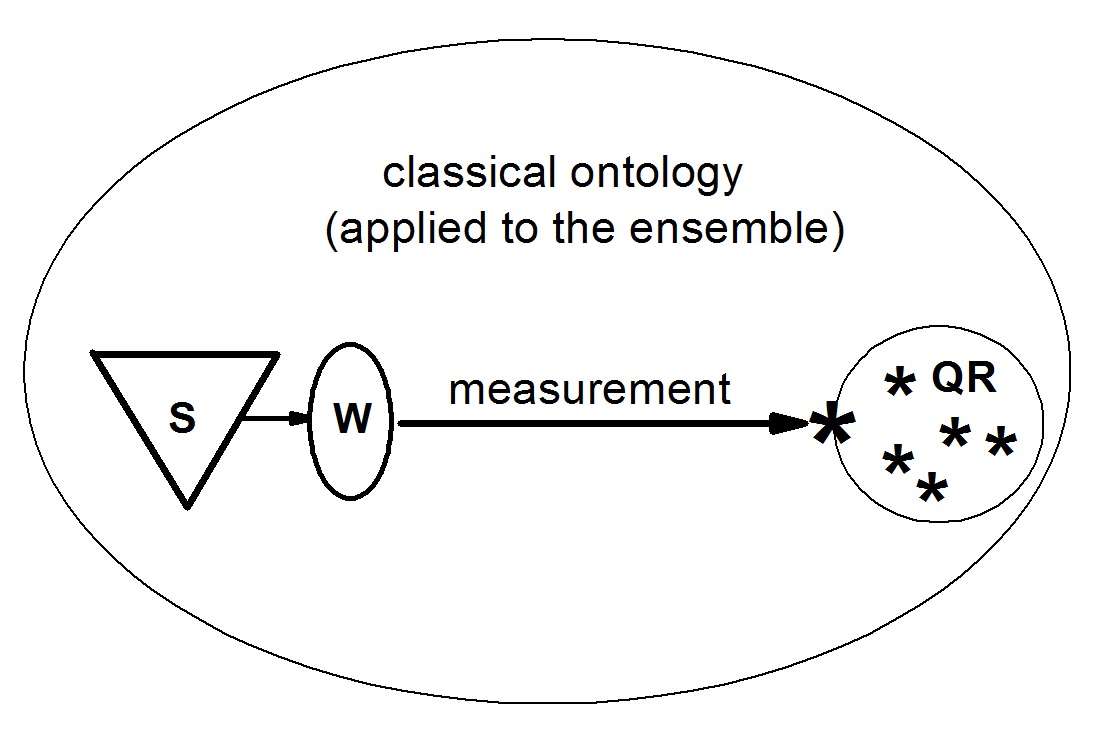}

Figure 8. Born-Einstein statistical interpretation, in which probabilities of a measurement result stem from a quantum state constituted by a kind of statistical {\it ensemble}.

After a vigorous discussion in the specialized literature, which also reached the public domain, Einstein and Max Born came to the conclusion that the quantum description is not applicable to an individual object, but to an ensemble of objects (not to be confused with Everett's interpretation). In this sense, the wavefunction contains information regarding the probabilities of measuring the allowable values of the physical properties of the objects in the ensemble [11] (Fig. 8).

This way the probabilistic nature of the quantum predictions can be understood. The interpretation is considered minimal as far as the assumptions are made, and constitutes a zero-order framework for solving the known problems of QM. In other words, it is suggested that QM is a kind statistical theory, and there is room for a more pointed description of the quantum Reality in the future.

\subsection{Instrumentalism}

\includegraphics[scale=0.45]{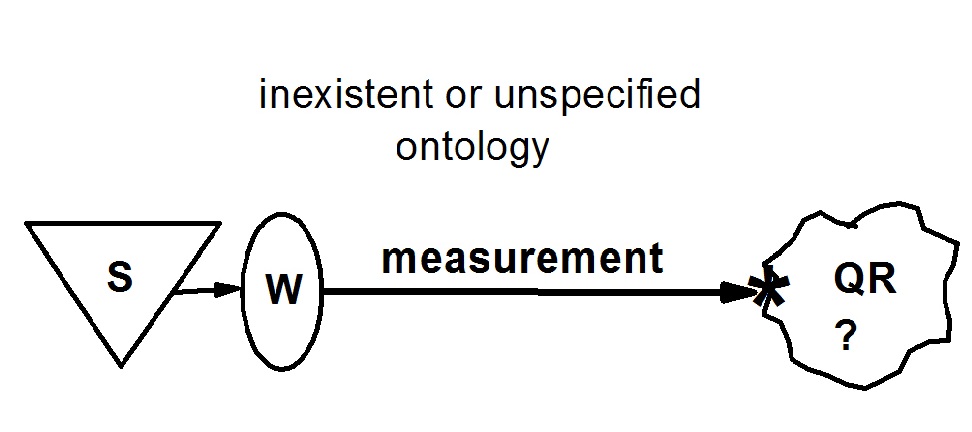}

Figure 9. The Instrumentalism interpretation, in which no QR is of concern.

In this interpretation the quantum description is a set of calculation rules, without any ontological pretension, that is, it explicitly renounces the logos of Quantum Reality and focuses just on the language that connects the Subject $S$ with the results of experiments. Its domain is the Wittgenstein filter set (now strongly focused on the mathematical formalism). In some sense, this interpretation is a natural evolution of the early Copenhagen interpretation, as noted above (Fig. 9).

\subsection{The Quanton interpretation}

\includegraphics[scale=0.4]{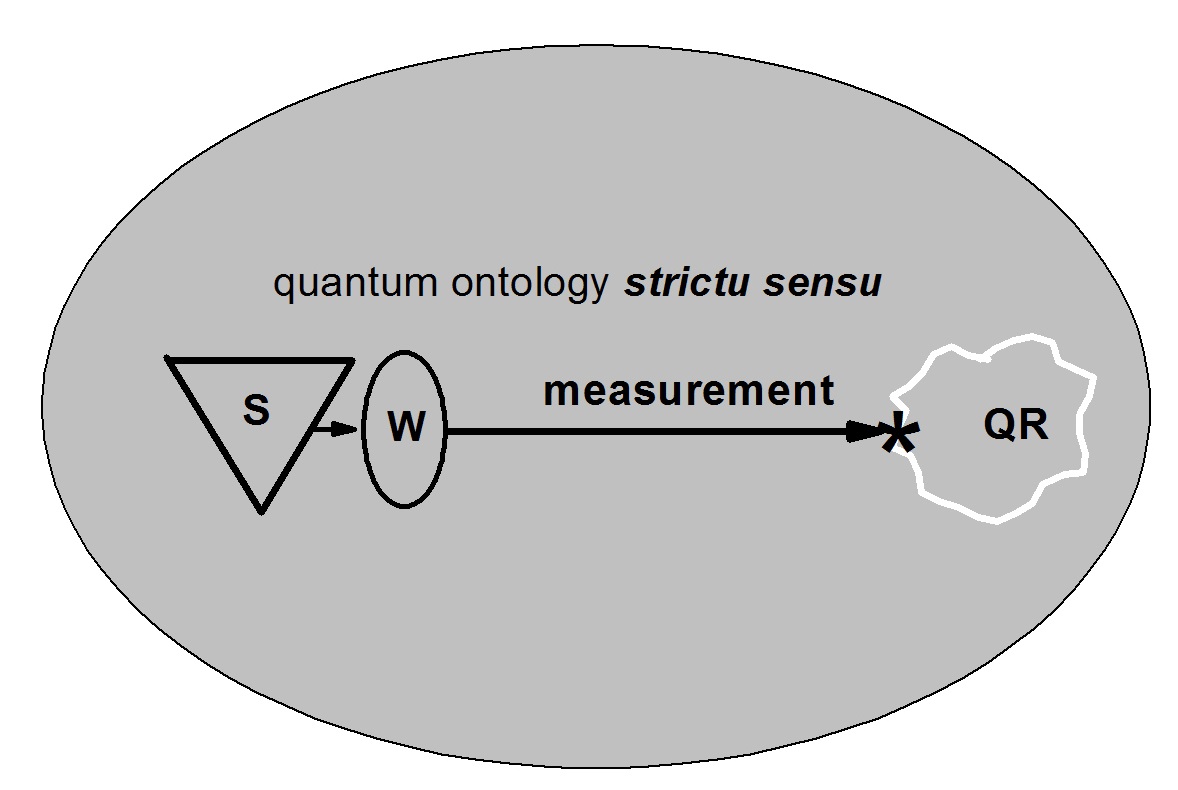}

Figure 10. The Quanton interpretation assumes that quantum objects are very different than classical ones from scratch, and the classical limit applies to the macroworld.

This interpretation builds on the proposal of a {\it quanton}, a term coined by Bunge [22]. The nature of objects is supposed to be different from the classic picture: they do not have definite values (but they do exist!) for any instant of time. The quantum formalism allows to calculate the expected values for a measurement (Fig. 10). But there is nothing idealistic or mystical about this, it is just the correction of the extrapolation made since the early days from the classical to the quantum world, carried out on the ontological level. In other words, since we only know classical objects in a direct unambiguous fashion, we have attributed to quantum objects similar properties which turn out to be misleading: it makes no sense, for example, to talk about a wave-particle duality, because these are classical views not to be applied to the quantum world objects, which do not possess such qualities as we know them. Electrons and not waves, nor small balls, but quite {\it sui generis} entities.

\section{Conclusions}

We have argued in this work for a more open and multiple representation of one of the most difficult issues of the 20th century Physics, Quantum Mechanics, a revolutionary theory that still hides its true meaning for practitioners and educators/students. The interpretation of QM issue is a long-term one and is not likely to end soon.

The use of diagrams and graphs in Science is not new, as we pointed out in the Introduction. However, in the form presented above, the diagrams are close to schemes or tools for identifying the various possible interpretations of QM and, consequently, the specific barriers of its effective knowledge and meaning for each case.

Therefore, as a useful tool, we constructed a set of Conceptual Diagrams for QM Interpretations, flexible enough to be applied to any interpretation of Quantum Mechanics, quite related in their essence to Venn's diagram of set theory. These have been shown explicitly for a (quite trivial) Classical Physics case and nine popular interpretations of QM, although there are many more proposals available which have not been addressed here [23]. Although the diagrams do not immediately solve any problem, their merging with the rest of cognitive tools may prove important in the long run. In many senses they could be helpful as the axiomatization of QM is (see, for example, [24], a tool to understand the inconsistencies and fundamentals of the theory related to its many interpretations put forward to make physical sense of it.

We are aware that this initial proposal must be examined in-depth, refined and extended. Since understanding relies on diagrams as elements entangled with mathematics, verbal language, graphics, and other tools, we believe the issue of QM and its difficulties can be better grasped with the aid of the presented Conceptual Diagrams. In the long run, it is hoped that they may be merged with other aspects of QM exposition and study, following the path of previous examples. Besides helping philosophers and physicists, we believe that this tool would be particularly useful for education and teaching.

\section*{Acknowledgments}

This work was performed under the auspices of a Research Fellowship granted by the {\it CNPq Agency}, Brazil and {\it FAPESP Agency}, S\~ao Paulo State through the grant {2020/08518-2}. 

\nocite{*}

\bigskip
\section*{References}

1] Lemke, ~J.~L., Multiplying Meaning: Visual and Verbal Semiotics in Scientific Text (Routledge, London), In: J.R. Martin and R. Veel (eds.) Reading Science. (1998)

\bigskip
\noindent
2] Lemke, ~J.~L., Talking Science: Content, Conflict, and Semantics. ERIC Documents Service (ED 282 402), Arlington. Paper presented at the American Educational Research Association Meeting, Washington DC, 1987

\bigskip
\noindent
3] Edwards,~A.~W.~F., Cogwheels of the Mind: The Story of Venn Diagrams. Johns Hopkins University Press, Baltimore, 2004

\bigskip
\noindent
4] Kuhn, ~T., The Structure of Scientific Revolutions. University of Chicago Press, Chicago, 1996

\bigskip
\noindent
5] 't Hooft and Veltman, ~M., DIAGRAMMAR, CERN Report73/9, reprinted in Particle Interactions at Very High Energies. Nato Adv. Study Inst.Series, Sect. B 4b, 177, 1973

\bigskip
\noindent
6] Latour, ~B., Visualisation and Cognition: Drawing Things Together. Jai Press, Stanford. In: H. Kuklick (ed.) Knowledge and Society Studies in the Sociology of Culture Past and Present, 1986 

\bigskip
\noindent
7] Airey, J., Science, Language and Literacy: Case Studies of Learning in Swedish University Physics. PhD Thesis (Uppsala U., Sweden) 2009

\bigskip
\noindent
8] de la Pe\~na,~L. Introduction to Quantum Mechanics. Ediciones Cient\'\i ficas Universitarias, M\' exico, 2010

\bigskip
\noindent 
9] Schmek,~A. and Fischer,~H.~E., Multiple representations in Physics and Science Education - Why should we use them? Springer Models and Modelling in Science Education, vol. Z, Heidelberg 2017 

\bigskip
\noindent
10] Heisenberg,~W. In: https://www.aip.org/history-programs/niels-bohr-library/oral-histories/4661-8. Oral Histories, AIP, NY, 1963
\bigskip
\noindent 
11] {Ismael, ~J. In: https://plato.stanford.edu/archives/fall2021/entries/qm/}, Quantum Mechanics. The Stanford Encyclopedia of Philosophy (Fall 2021 Edition), Edward N. Zalta (ed.), 2021

\bigskip
\noindent 
12] Bunge, ~M., Physics and Philosophy. Ed. Perspectiva, S\~ao Paulo, 1992

\bigskip
\noindent
13] Wittgenstein, L., Philosophical Investigations. Mc Millan, NY, 1953

\bigskip
\noindent 
14] Wittgenstein, L., Tractatus Logico-Philosophicus. Routledge, NY, 1994

\bigskip
\noindent
15] Petersen,~A., The philosophy of Niels Bohr. Bull. Atomic Scientists 19, 8 (1963)

\bigskip
\noindent
16] Bunge, M., Twenty-Five Centuries of Quantum Physics: From Pythagoras to Us, and from Subjectivism to Realism. Science \& Education, 12, 445 (2003)

\bigskip
\noindent
17] Herbert,~N. Quantum Reality. Anchor Books, NY, 1985

\bigskip
\noindent
18] Pinto Neto,~A., Theories and Interpretations of Quantum Mechanics. Ed. Livraria da F\'\i sica, S\~ao Paulo, 2010

\bigskip
\noindent
19] Borges, ~J.~L., The garden of forking paths. Penguin Books, UK, 2018

\bigskip
\noindent
20] Merrel,~F., Unthinking Thinking: Jorge Luis Borges, Mathematics, and the New Physics. Purdue University Press, West Lafayette, 1991

\bigskip
\noindent  
21] Anonymous, Hindu Myths: A Sourcebook Translated from the Sanskrit. Penguin Books, UK, 2004

\bigskip
\noindent
22] Bunge, ~M., Philosophy of Physics. Reidel, Dordretch, 1973

\bigskip
\noindent 
23] Johansson,~L.-G., Interpreting Quantum Mechanics. Taylor \& Francis, UK, 2007

\bigskip
\noindent
24] P\'erez Bergliaffa,~S.~E., Romero,~G.~E. and Vucetich,~H., Axiomatic foundations of quantum mechanics revisited: the case for systems. Int. J. Theor. Phys. 35, 1805 (1995)

\end{document}